\documentclass[reqno]{amsart}
\usepackage{amsmath}
\usepackage{amssymb}
\usepackage{amsfonts}
\usepackage{mathrsfs}
\usepackage[all,arc]{xy}
\usepackage{enumerate}
\usepackage{mathrsfs}
\usepackage{amsthm}
\usepackage{enumitem}
\usepackage{extarrows}
\usepackage{verbatim}
\usepackage{dsfont}
\usepackage{xcolor}
\usepackage{textcomp}
\usepackage{tikz-cd}
\usepackage{smartdiagram}
\usepackage{tensor}
\usepackage{mathtools}
\usepackage{float}
\usepackage[toc,page]{appendix}
\usepackage{enumitem}
\usetikzlibrary{decorations.pathmorphing}
\usetikzlibrary{backgrounds, chains, positioning, shadows}

\theoremstyle{definition}

\newtheorem{obs}{Observation}
\newtheorem{rem}{rem}

\makeatletter

\newcommand*{\rom}[1]{\expandafter\@slowromancap\romannumeral #1@}

\makeatletter
\begin{document} 
	\title{On the suitability of generalized regression neural networks for GNSS position time series prediction for geodetic applications in geodesy and geophysics}   
	\author[M. Kiani]{M. Kiani$^\dag$\\
		$^\dag$ \tiny School of surveying and geospatial data engineering, University of Tehran, Tehran, Iran}
	
	\thanks{Corresponding author, email:mostafakiani@ut.ac.ir, tel:+989100035865}
	\date{}
\begin{abstract}
	In this paper, the generalized regression neural network is used to predict the GNSS position time series. Using the IGS 24-hour final solution data for Bad Hamburg permanent GNSS station in Germany, it is shown that the larger the training of the network, the higher the accuracy is, regardless of the time span of the time series. In order to analyze the performance of the neural network in various conditions, 14 permanent stations are used in different countries, namely, Spain, France, Romania, Poland, Russian Federation, United Kingdom, Czech Republic, Sweden, Ukraine, Italy, Finland, Slovak Republic, Cyprus, and Greece. The performance analysis is divided into two parts, continuous data-without gaps-and discontinuous ones-having intervals of gaps with no data available. Three measure of error are presented, namely, symmetric mean absolute percentage error, standard deviation, and mean of absolute errors. It is shown that for discontinuous data the position can be predicted with an accuracy of up to 6 centimeters, while the continuous data positions present a higher prediction accuracy, as high as 3 centimeters. In order to compare the results of this machine learning algorithm with the traditional statistical approaches, the Theta method is used, which is well-established for high-accuracy time series prediction. The comparison shows that the generalized regression neural network machine learning algorithm presents better accuracy than the Theta method, possibly up to 250 times. In addition, it is approximately 4.6 times faster.   
\end{abstract}
\maketitle
$Key words:$ GNSS position time series, generalized regression neural networks, machine learning, historical data, training data, prediction accuracy, symmetric mean absolute percentage error, standard deviation, mean of absolute errors
\section{Introduction}
Neural networks and machine learning algorithms have been established as powerful and promising methods that have widespread applications in various fields, such as economics \cite{Makridakis}, fluid dynamics \cite{Liu}, stock market \cite{Makridakis,Dash} and alike. These methods can be used for different purposes, including classification, prediction, the so-called regime switching, and asynchronous sampling \cite{Orr}. One of the most interesting purposes of these methods is the prediction of future outcomes. This is usually done using a set of historical data, called time series. In this application, the input layer of the neural network is the historical data, hidden layer a set of mathematical functions used for approximation, and the output layer the prediction of the next values of the time series \cite{Makridakis}. The choice of the mathematical functions in the hidden layer is what distinguishes different neural networks from each other \cite{Orr,Makridakis}. Some of the most widely known neural networks in different applications include Bayesian \cite{Lauret}, Radial Basis Functions (RBF) \cite{Zhang}, multilayer perceptron \cite{Boughrara}, k-nearest neighbor \cite{Rajini}, CART regression trees \cite{Hayes}, support vector regression \cite{Awad}, Gaussian processes \cite{Mathur}, recurrent neural network \cite{Du}, long short term memory \cite{Fischer}, and generalized regression \cite{Li}.

In the field of earth science, and more specifically the field of satellite geodesy, the long-term changes in the environment and the earth's processes are of much interest. In order to perform such analyses, one needs to have a set of historical data in the form of time series. Many applications can then be introduced based on the time series. For instance, the direction and velocity of the earth's plates movements are investigated \cite{Ito}. Another application of geodetic time series is the prediction of the future of the time series. This application can be of great importance in the position time series, because changes in the position of a station may have different and sometimes serious implications, such as subsidence of the ground in which the station is located, or even tremors of an earthquake. Hence, it seems necessary to investigate different methods of prediction. One approach is using the traditional statistical methods such as Theta \cite{Assimakopoulos}. However, in order to achieve better accuracies in prediction, we need to use other methods. This is the motivation of the present paper.

One of the promising methods for time series prediction is neural network or machine learning. The mentioned methods in the first paragraph are all used for prediction of time series in different fields. However, there has been little practical literature on the use of these methods in geodetic time series \cite{Kiani,Fazilova}. In the field of satellite geodesy, the time series are in the form of position or residual position, gathered by receivers from the GNSS. Like other fields, various machine learning algorithms can be used for prediction. \cite{Kiani} has shown that the RBF are probably not a good choice for predicting the residual time series. Thus, other methods must be evaluated. On of the methods is the Generalized Regression Neural Networks (GRNN). The present paper is motivated by the following characteristics of GRNN, to prefer it over other methods:
\\

$\bullet$ Its simplicity; such that unlike other methods such as RBF, it does not require matrix handlings (multiplication, inversion, and regularization) \cite{Zhang,Kiani}.

$\bullet$ Its high speed; such that it facilitates the time-consuming neural network processes.

$\bullet$ Its high accuracy; which in many cases outperforms other neural networks \cite{Makridakis,Ahmed}.\\

As it will be shown at the end of this paper, this choice of neural networks is quite effective and promising, such that it can reach up to 3 centimeters in prediction accuracy.

The major contributions of the present paper can be summarized as the following\\

$\bullet$ Analysis of the effect of the training-size-the number of historical data used for the prediction-on the accuracy of prediction.

$\bullet$ Analysis of the effect of the discontinuity of data on the accuracy of prediction.

$\bullet$ Analysis of the prediction results in relatively large number of time series in different countries (with different environmental conditions and types of receivers), using different accuracy evaluation criteria.
\\

In the following sections the materials are presented as follows. In section 2, the mathematical background of the GRNN is presented. Section 3 deals with the application of GRNN in GNSS position time series prediction. Finally, section 4 is devoted to the conclusions and suggestions for future works.   

\section{Mathematical background}
In this section, the essential backgrounds for the GRNN are presented. 

The method of GRNN, introduced by the seminal papers \cite{Nadaraya,Watson}, considers the prediction value as a linear composition of the training values. Like RBF neural networks, there is no weight between the input and hidden layers. If $x_i,~i=1,...,p$ are the input nodes, $\omega_j,~j=1,...,v$ the weights between the hidden and the output layer, and $y_i,~i=1,...,p$ the output nodes, then the prediction for the $k$th output, denoted by $\hat{y}_k$ and related to the input $x_k$, is given in the following form
\begin{equation}\label{eqn1}
\hat{y}_k=\sum_{l=1}^{v}\omega_l z_l, 
\end{equation}
in which $z_l \in Z=\{\hat{y}_s,\hat{y}_{s+1},...,\hat{y}_{v-1}\}$. $Z$ denotes the training set and in each iteration it changes, since the change $\hat{y}_{v-1}=\hat{y}_k$ is made. The value of $v$ determines how many of the data are used for the training and is normally called training-size. 
\begin{obs}\label{obs1}
	Note that the number of all input data is $p$. The training size-number of point used for prediction-is $v$. Number of outputs is, of course, $p$. 
\end{obs}

One advantage of the GRNN is that the weights $\omega_j,~j=1,...,q$ are known analytically, which implies that there is no need for computationally expensive matrix operations as in RBF neural networks. These values are given in the following form \cite{Ahmed}
\begin{equation}\label{eqn2}
\omega_j=\frac{K(\frac{||x-x_j||}{h})}{\sum_{l=1}^{v}K(\frac{||x-x_l||}{h})},\quad j=1,...,v, 
\end{equation}
where $h$ is the band parameter and $K$ is the kernel function. 
\begin{rem}\label{rem1}
	Note that although there are different choices for the kernel function, the Gaussian kernel is particularly important and has been used widely for the prediction purposes. For this reason, we employ this function for the time series prediction, and as it will be seen in the next chapter, this choice is quite effective. The Gaussian kernel has the following form
	\begin{equation}\label{eqn3}
	K(A)=\frac{e^{-\frac{A^2}{2}}}{\sqrt{2\pi}}.
	\end{equation}
\end{rem}

After computing $\hat{y}_k$ in Eqn. \eqref{eqn1}, the error of the prediction, $E$, can be computed based on the real value $y_k$, as the following
\begin{equation}\label{eqn4}
E=y_k-\hat{y}_k.
\end{equation}
The absolute value of $E$ is a criterion that determines the need for further processes. If $|E|<T$, with $T$ being a given threshold, the prediction is acceptable. However, the number of training data, i.e. the training-size must increase.

With these preliminaries, the GRNN process can be summarized in the following diagram

\begin{center}
	\resizebox{0.6\linewidth}{!}{%
		\begin{tikzpicture}[
		node distance = 10mm and 0mm,
		start chain = going below,
		box/.style = {rectangle, rounded corners, draw=gray, very thick,
			minimum height=16mm, text width=60mm, align=flush center,
			top color=#1!90, bottom color=#1!10,
			drop shadow, on chain},
		down arrow/.style = {
			single arrow, draw,
			minimum height=2.5em,
			transform shape,
			rotate=-90,
		}
		]
		\node (n1) [box=cyan]{$v<p$? then choose the training-size $v$ to be used in Eqn. \eqref{eqn1} and Eqn. \eqref{eqn2}};
		\node (n2) [box=cyan]{compute $\hat{y}_k$ in Eqn. \eqref{eqn1}};
		\node (n3) [box=cyan]{compute the absolute value of prediction error in Eqn. \eqref{eqn4}};
		\node (n4) [box=cyan]{compare the absolute value of error, $|E|$ with the given threshold $T$};
		\node (n5) [box=cyan]{$|E|$<T?};
		\node (n6) [box=cyan]{finish; the prediction is accurate enough. Go to the next prediction};
		\draw [black, thick, ->] (n2) edge (n3) (n3) edge (n4)(n4) edge (n5);
		\path (n1) -- node (yes) {\emph{yes}} (n2);
		\draw[green, thick, ->] (n1) -- (yes) -- (n2);
		\path (n1.west) -- ++(-40pt,0pt) node (no) {\emph{no}} (n6.west);
		\draw[red, thick, ->] (n1) -- (no) |-(n6);
		
		\path (n5) -- node (yes) {\emph{yes}} (n6);
		\draw[green, thick, ->] (n5) -- (yes) -- (n6);
		\path (n5.east) -- ++(40pt,0pt) node (no) {\emph{no}} (n1);
		\draw[red, thick, ->] (n5) -- (no) |-(n1);
		
		\end{tikzpicture}
	}
\end{center}
It is important to notice that the overall accuracy of the GRNN method can be assessed with different criteria. In this paper, we have used three different accuracy criteria, namely, symmetric Mean Absolute Percentage Error (sMAPE) \cite{Makridakis,Ahmed}, Standard Deviation (StD), and Mean of Absolute Errors (MAbs). These criteria are defined as the following
\begin{equation}\label{eqn5}
\begin{split}
&sMAPE=\sum_{i=1}^{N}\frac{|y_i-\hat{y}_i|}{|y_k|+|\hat{y}_k|},\\
&StD=\sqrt{\sum_{i=1}^{N}\frac{|h_i-\bar{h}|^2}{N-1}},\\
&MAbs=\sum_{i=1}^{N}\frac{|y_i-\hat{y}_i|}{N},
\end{split}
\end{equation}  
in which $N$ denotes the total number of predicted values (those values used for training are not used), $h_i=(y_i-\hat{y}_i)$, and $\bar{h}=(\sum_{i=1}^{N}\frac{h_i}{M})$. Note that sMAPE is expressed in terms of percentage, while the other two have the same unit as that of time series (usually meter).
\begin{rem}\label{rem2}
	All of the three mentioned criteria represent a similar meaning: the smaller the error criterion, the more accurate the prediction is. However, sMAPE is a measure of how the errors correspond to the time series values, i.e. relative accuracy. StD is interpreted as a measure of data dispersion. The MAbs represents bias of prediction with respect to the mean value of predicted values.
\end{rem}
\section{Application of GRNN in GNSS position time series prediction}
\begin{figure}[H]
	\centering
	\includegraphics[width=1\linewidth]{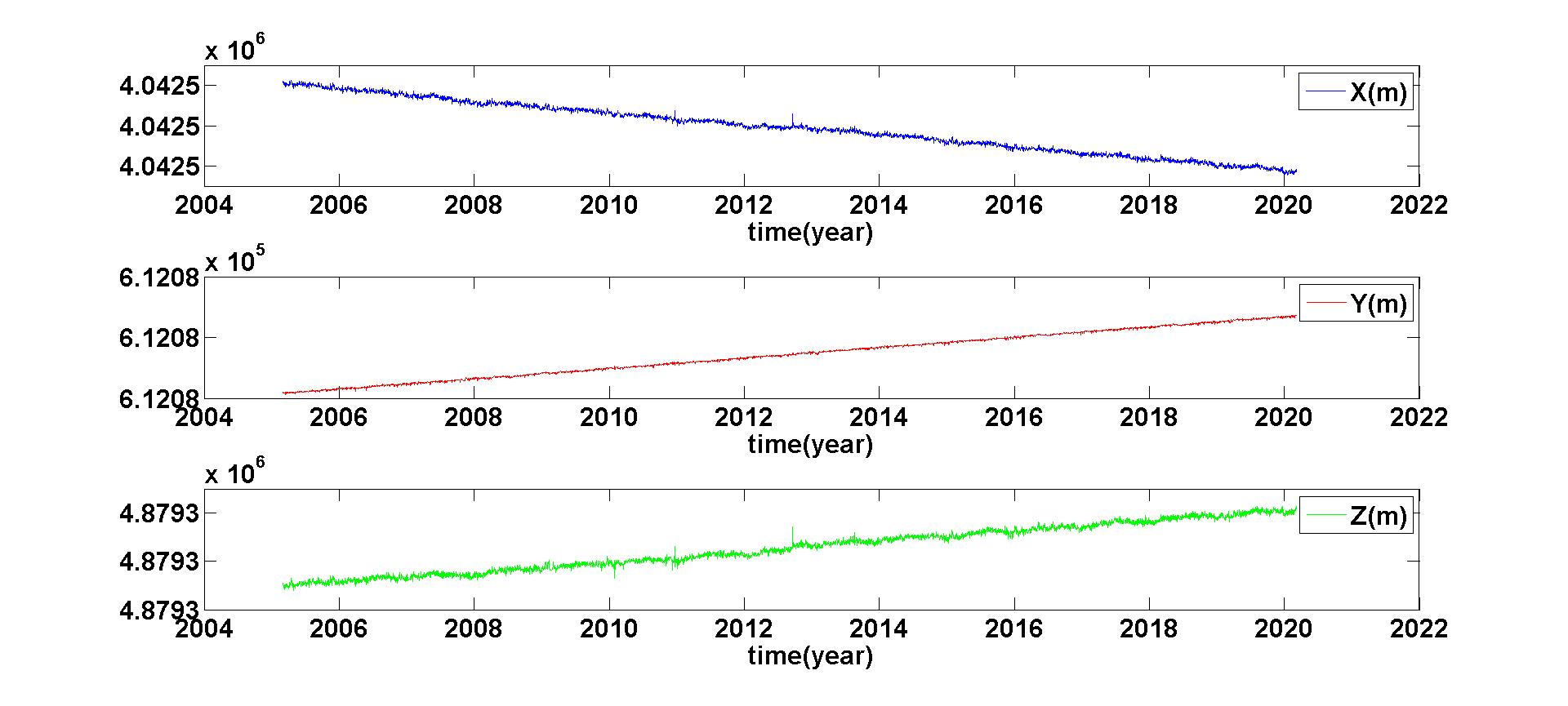}
	\caption{Bad Hamburg permanent GNSS station position time series, in IGS14 system, spanning from March 2005 to March 2020}
	\label{fig1}
\end{figure}

\begin{figure}[H]
	\centering
	\includegraphics[width=1\linewidth]{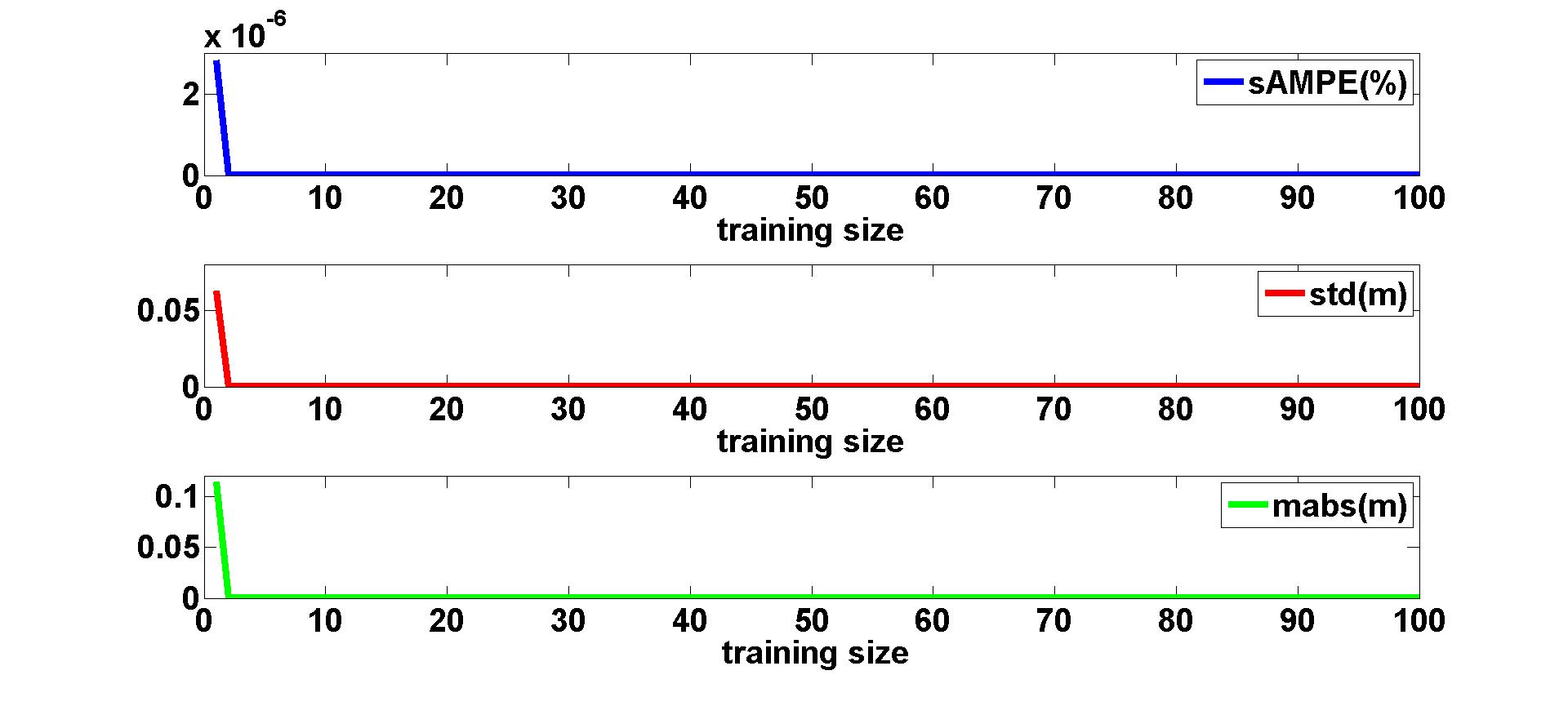}
	\caption{Effect of the training-size on the prediction accuracy criteria, $X$ component of the time series under consideration}
	\label{fig2}
\end{figure}

\begin{figure}[H]
	\centering
	\includegraphics[width=1\linewidth]{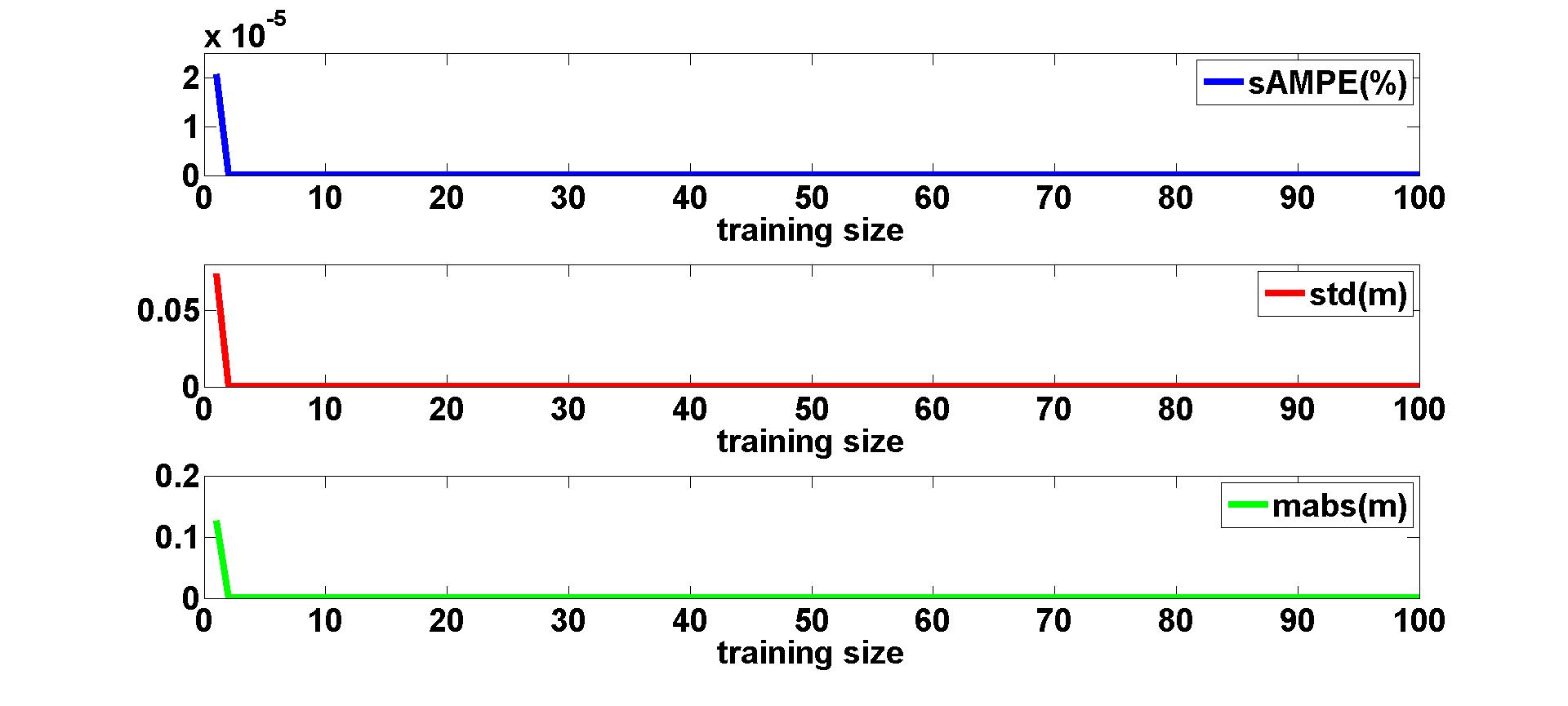}
	\caption{Effect of the training-size on the prediction accuracy criteria, $Y$ component of the time series under consideration}
	\label{fig3}
\end{figure}

\begin{figure}[H]
	\centering
	\includegraphics[width=1\linewidth]{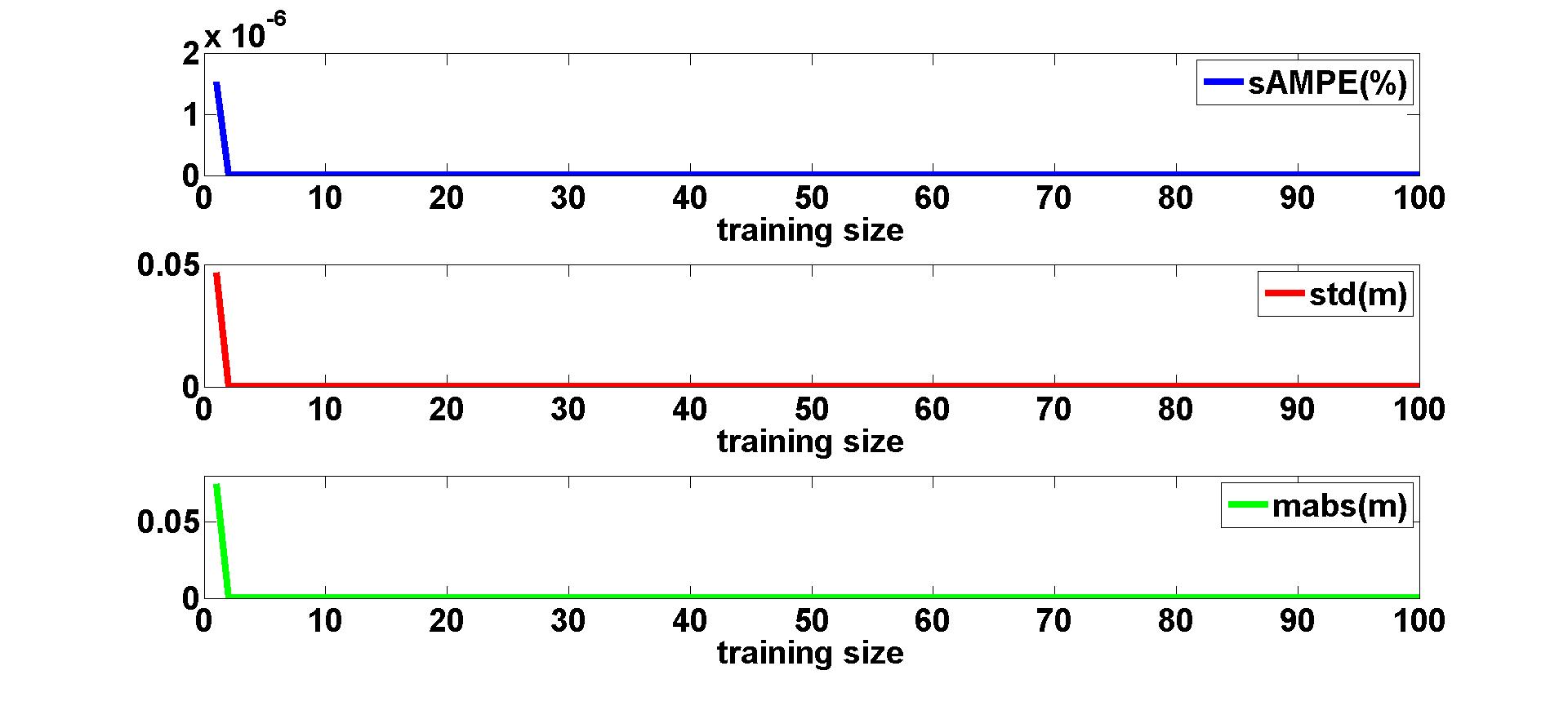}
	\caption{Effect of the training-size on the prediction accuracy criteria, $Z$ component of the time series under consideration}
	\label{fig4}
\end{figure}

\begin{figure}[H]
	\centering
	\includegraphics[width=1\linewidth]{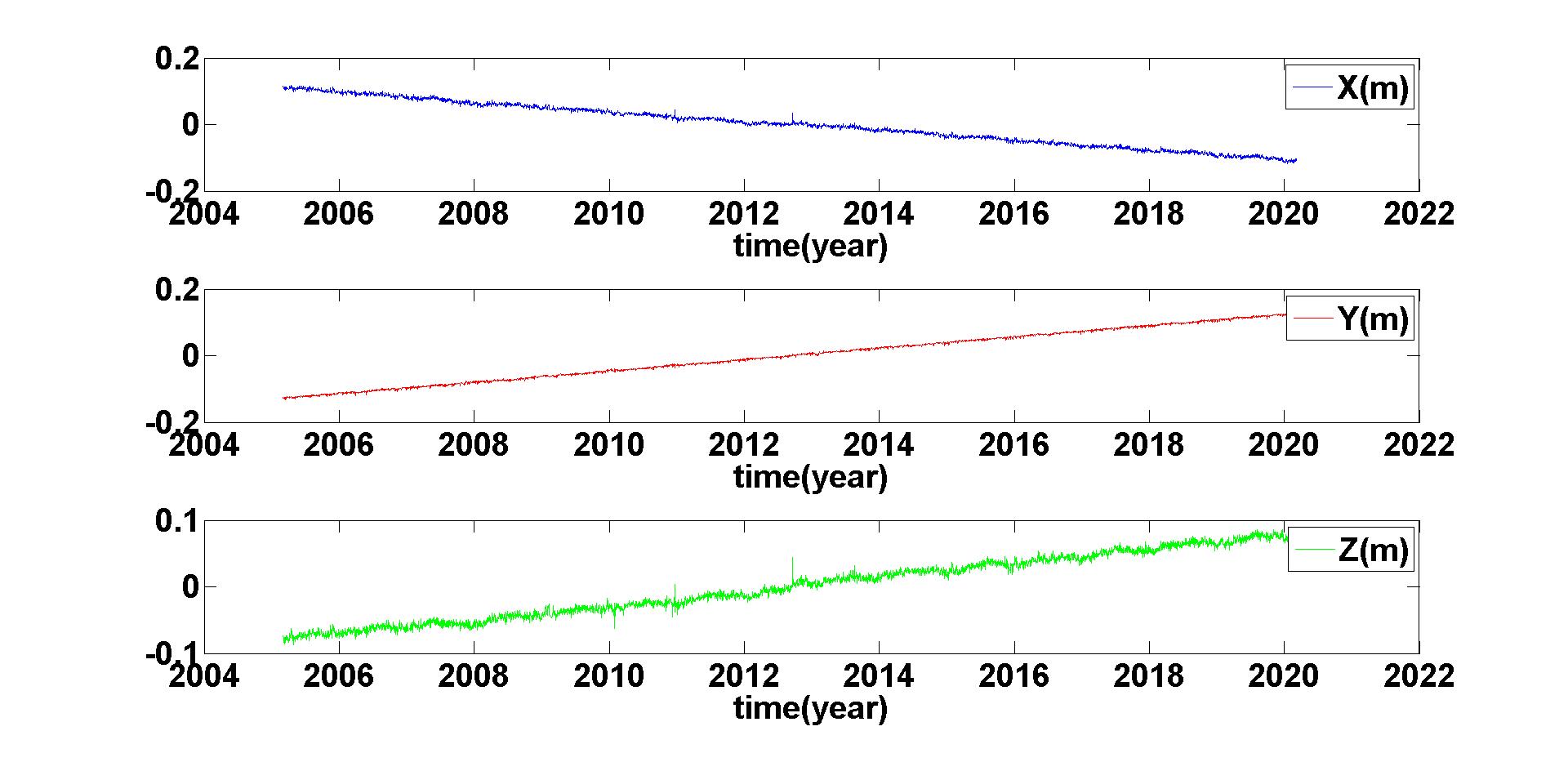}
	\caption{The change in coordinates of the Bad Hamburg station with respect to the mean value of the components}
	\label{fig5}
\end{figure}

In this section, the application of GRNN is presented in the context of GNSS position time series prediction. The time series for this study are taken from the IGS 24-hour final solution data \cite{Blewitt}. The time series represents the three-dimensional coordinates of the permanent station for years. First, we intend to analyze the effect of training-size on the accuracy of prediction. For this reason, we have used the data from Bad Hamburg permanent GNSS station. The original time series is shown in Fig. \ref{fig1}. The coordinates are in the IGS14 (ITRS) system.
The effect of training-size is shown in Fig. \ref{fig2}, \ref{fig3}, and \ref{fig4}. In these figures, the training-size is increased from 1 to 100, and sMAPE, StD, and MAbs are computed for each of the three components $X$, $Y$, $Z$.   
It can be understood from Fig. \ref{fig1}, \ref{fig2}, and \ref{fig3} that in GRNN, the training-size has a positive impact on the accuracy of the prediction, unlike the RBF \cite{Kiani}. Besides, the rate of improvement in the accuracy is quite good, such that increasing the training-size from 1 to 2, reduces sMAPE with a scale of 4.7$\times 10^{-6}$, StD with a scale of 5$\times 10^{-6}$, and MAbs with a scale of 4.7$\times 10^{-6}$. In the training-size ($v=$100), sMAPE, StD, and MAbs, are, respectively, 3.2$\times 10^{-13}\textdiscount$, 7.9$\times 10^{-9}$, and 1.3$\times 10^{-8}$. This is excellent indeed. The results imply that the next outcomes of the time series can be predicted with an accuracy of nanometer, in terms of StD. However, the time series presented in Bad Hamburg is relatively smooth and does not have any anomalies in the training data. Besides, the amplitude of the change of $X, Y, Z$ components, with respect to their corresponding mean values, are small (around 30 centimeters in 15 years for $X, Y$, and 20 centimeters in 15 years for $Z$, see Fig. \ref{fig5}).
\subsection{Prediction accuracy of GRNN in various time series in different countries}  
Note that the time series in Fig. \ref{fig1}. is continuous, meaning it has no gaps in the data. Sometimes, because of various reasons such as the unavailability of the receivers or their failure, the data are discontinuous, meaning they have gaps. We are interested in the effect of gaps on the accuracy of prediction. In order to analyze the effect of the amplitude of changes and discontinuity of data on the accuracy of prediction, 14 permanent GNSS stations in 14 different countries are used, which include both continuous and discontinuous data, and also both the little changing and highly changing time series. The results of applying the GRNN algorithm on these time series are shown in tables 1, 2, and 3.  
\begin{table}[H]
	\label{table1}
	\centering
	\caption{Prediction accuracy of the GRNN for 14 different permanent GNSS stations, $X$ component}
	\begin{tabular}{|c|c|c|c|c|c|}\hline
		\cline{1-6}station name & time span (year) & state of data & sMAPE$\textdiscount$ & StD (m) & MAbs(m)\\
		
		\cline{1-6} A Coruna, Spain & 1998-2020 & \textcolor{red}{discontinuous} & 6$\times 10^{-4}$ & 14.383 & 27.706\\
		
		\cline{1-6} Ajaccio, France & 2000-2020 & \textcolor{red}{discontinuous} & 8$\times 10^{-5}$ & 2.859 & 3.858\\
		
		\cline{1-6} Bacau, Romania & 2006-2020 & continuous & 3$\times 10^{-6}$ & 0.189 & 0.141\\
		
		\cline{1-6} Borowa Gora, Poland & 1996-2020 & continuous & 4$\times 10^{-5}$ & 1.255 & 1.636\\
		
		\cline{1-6} Svetloe, Russian Federation & 1997-2020 & continuous & 4$\times 10^{-4}$ & 2.496 & 10.906\\
		
		\cline{1-6} Morpeth, United Kingdom & 1996-2020 & \textcolor{red}{discontinuous} & 3$\times 10^{-4}$ & 3.539 & 13.444\\
		
		\cline{1-6} Kunzak, Czech Republic & 2005-2020 & continuous & 3$\times 10^{-6}$ &  0.132 & 0.130\\
		
		\cline{1-6} Maartsbo, Sweden & 1996-2020 & continuous & 1$\times 10^{-5}$ & 0.154 & 0.499\\
		
		\cline{1-6} Mariupol, Ukraine & 2013-2020 & continuous & 2$\times 10^{-6}$ & 0.073 & 0.099\\
		
		\cline{1-6} Matera, Italy & 1994-2020 & continuous & 3$\times 10^{-5}$ & 0.727 & 1.423\\
		
		\cline{1-6} Kirkkonummi, Finland & 2013-2020 & continuous & 4$\times 10^{-5}$ & 2.027 & 1.356\\
		
		\cline{1-6} Modra-Piesok, Slovak Republic & 2007-2020 & continuous & 8$\times 10^{-6}$ & 0.247 & 0.347\\
		
		\cline{1-6} Nicosia, Cyprus & 1997-2020 & \textcolor{red}{discontinuous} & 2$\times 10^{-4}$ & 7.947 & 11.506\\
		
		\cline{1-6} Athens, Greece & 2006-2020 & continuous & 2$\times 10^{-6}$ & 0.153 & 0.125\\
		
		\hline
	\end{tabular}
\end{table}

\begin{table}[H]
	\label{table2}
	\centering
	\caption{Prediction accuracy of the GRNN for 14 different permanent GNSS stations, $Y$ component}
	\begin{tabular}{|c|c|c|c|c|c|}\hline
		\cline{1-6}station name & time span (year) & state of data & sMAPE$\textdiscount$ & StD (m) & MAbs(m)\\
		
		\cline{1-6} A Coruna, Spain & 1998-2020 & \textcolor{red}{discontinuous} & 4$\times 10^{-4}$ & 1.520 & 2.955\\
		
		\cline{1-6} Ajaccio, France & 2000-2020 & \textcolor{red}{discontinuous} & 4$\times 10^{-5}$ & 0.246 & 0.289\\
		
		\cline{1-6} Bacau, Romania & 2006-2020 & continuous & 5$\times 10^{-6}$ & 0.123 & 0.100\\
		
		\cline{1-6} Borowa Gora, Poland & 1996-2020 & continuous & 2$\times 10^{-5}$ & 0.326 & 0.284\\
		
		\cline{1-6} Svetloe, Russian Federation & 1997-2020 & continuous & 4$\times 10^{-4}$ & 1.511 & 6.319\\
		
		\cline{1-6} Morpeth, United Kingdom & 1996-2020 & \textcolor{red}{discontinuous} & 8$\times 10^{-5}$ & 0.065 & 0.093\\
		
		\cline{1-6} Kunzak, Czech Republic & 2005-2020 & continuous & 8$\times 10^{-6}$ &  0.058 & 0.088\\
		
		\cline{1-6} Maartsbo, Sweden & 1996-2020 & continuous & 1$\times 10^{-5}$ & 0.107 & 0.099\\
		
		\cline{1-6} Mariupol, Ukraine & 2013-2020 & continuous & 1$\times 10^{-6}$ & 0.035 & 0.029\\
		
		\cline{1-6} Matera, Italy & 1994-2020 & continuous & 9$\times 10^{-6}$ & 0.107 & 0.129\\
		
		\cline{1-6} Kirkkonummi, Finland & 2013-2020 & continuous & 5$\times 10^{-5}$ & 1.044 & 0.700\\
		
		\cline{1-6} Modra-Piesok, Slovak Republic & 2007-2020 & continuous & 1$\times 10^{-5}$ & 0.111 & 0.221\\
		
		\cline{1-6} Nicosia, Cyprus & 2007-2020 & \textcolor{red}{discontinuous} & 1$\times 10^{-4}$ & 4.516 & 5.536\\
		
		\cline{1-6} Athens, Greece & 2006-2020 & continuous & 2$\times 10^{-6}$ & 0.066 & 0.052\\
		
		\hline
	\end{tabular}
\end{table}

\begin{table}[H]
	\label{table3}
	\centering
	\caption{Prediction accuracy of the GRNN for 14 different permanent GNSS stations, $Z$ component}
	\begin{tabular}{|c|c|c|c|c|c|}\hline
		\cline{1-6}station name & time span (year) & state of data & sMAPE$\textdiscount$ & StD (m) & MAbs(m)\\
		
		\cline{1-6} A Coruna, Spain & 1998-2020 & \textcolor{red}{discontinuous} & 6$\times 10^{-4}$ & 12.924 & 24.790\\
		
		\cline{1-6} Ajaccio, France & 2000-2020 & \textcolor{red}{discontinuous} & 7$\times 10^{-5}$ & 2.433 & 3.243\\
		
		\cline{1-6} Bacau, Romania & 2006-2020 & continuous & 2$\times 10^{-6}$ & 0.228 & 0.120\\
		
		\cline{1-6} Borowa Gora, Poland & 1996-2020 & continuous & 3$\times 10^{-5}$ & 1.189 & 1.775\\
		
		\cline{1-6} Svetloe, Russian Federation & 1997-2020& continuous & 5$\times 10^{-4}$ & 6.596 & 28.546\\
		
		\cline{1-6} Morpeth, United Kingdom & 1996-2020& \textcolor{red}{discontinuous} & 3$\times 10^{-4}$ & 4.702 & 17.700\\
		
		\cline{1-6} Kunzak, Czech Republic & 2005-2020 & continuous & 1$\times 10^{-6}$ &  0.115 & 0.085\\
		
		\cline{1-6} Maartsbo, Sweden & 1996-2020 & continuous & 8$\times 10^{-6}$ & 0.226 & 0.465\\
		
		\cline{1-6} Mariupol, Ukraine & 2013-2020 & continuous & 9$\times 10^{-7}$ & 0.053 & 0.043\\
		
		\cline{1-6} Matera, Italy & 1994-2020 & continuous & 2$\times 10^{-5}$ & 0.456 & 0.875\\
		
		\cline{1-6} Kirkkonummi, Finland & 2013-2020 & continuous & 3$\times 10^{-5}$ & 2.562 & 2.162\\
		
		\cline{1-6} Modra-Piesok, Slovak Republic & 2007-2020 & continuous & 1$\times 10^{-5}$ & 0.375 & 0.699\\
		
		\cline{1-6} Nicosia, Cyprus & 1997-2020 & \textcolor{red}{discontinuous} & 2$\times 10^{-4}$ & 7.042 & 9.070\\
		
		\cline{1-6} Athens, Greece & 2006-2020 & continuous & 2$\times 10^{-6}$ & 0.076 & 0.105\\
		
		\hline
	\end{tabular}
\end{table}

\subsection{Analysis of the results}
\begin{figure}[H]
	\centering
	\includegraphics[width=1\linewidth]{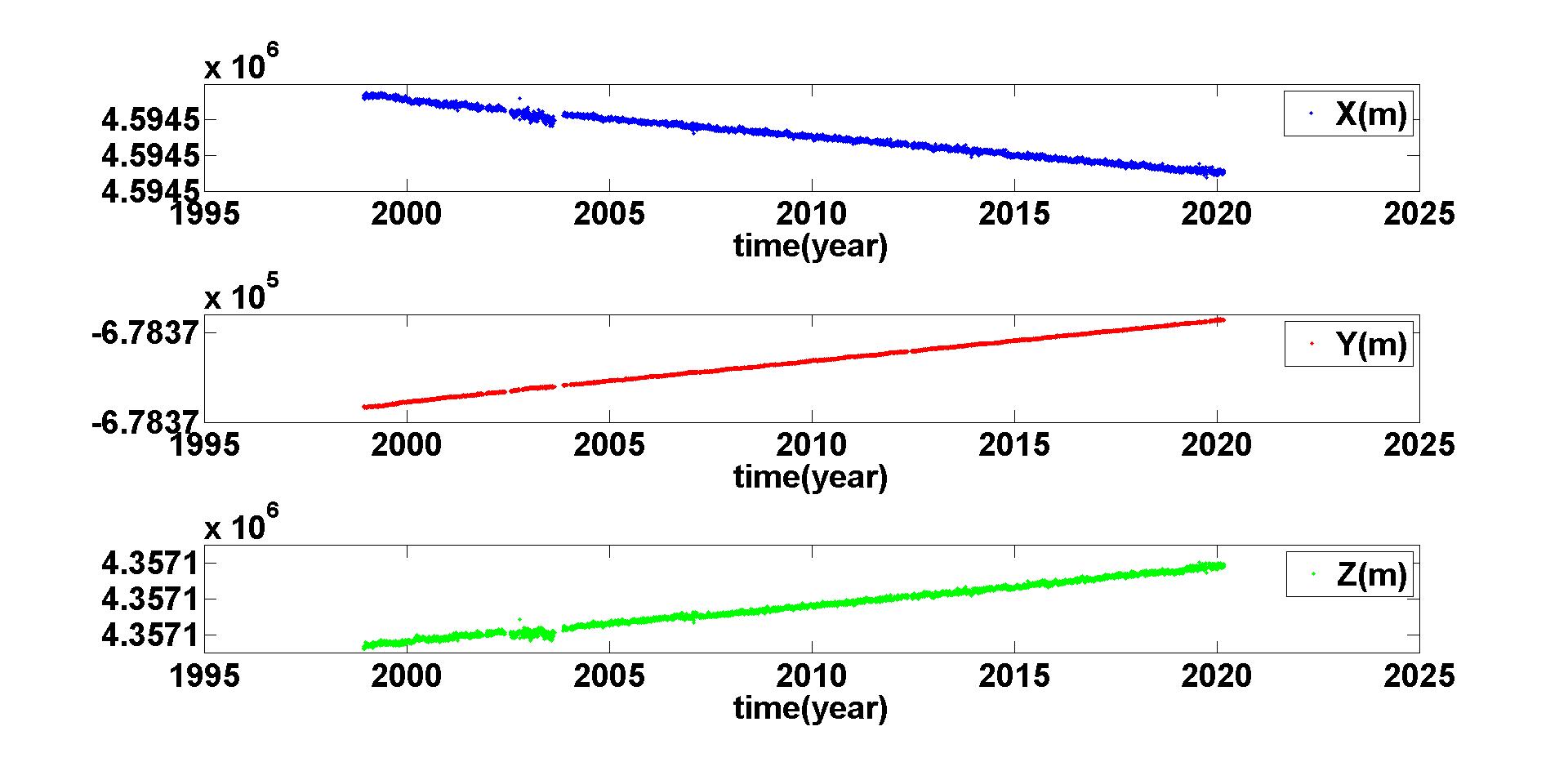}
	\caption{Discontinuous time series of A Coruna, Spain}
	\label{fig6}
\end{figure}

\begin{figure}[H]
	\centering
	\includegraphics[width=1\linewidth]{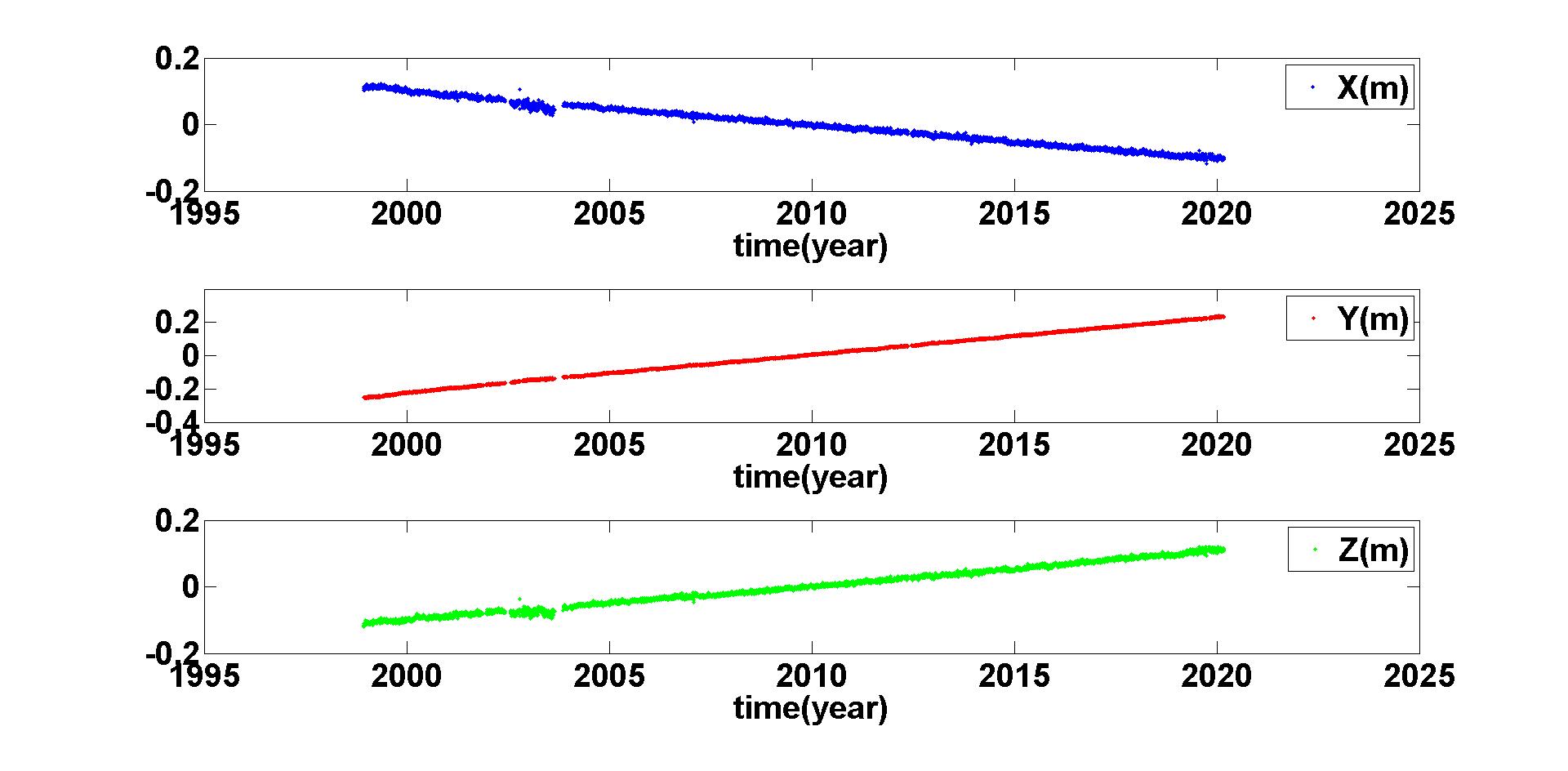}
	\caption{Amplitude of changes in the A Coruna, Spain time series, with respect to the mean value of components}
	\label{fig7}
\end{figure}

From the tables 1, 2, and 3, it can be concluded that the worst prediction accuracy is for the A Coruna, Spain GNSS station. This is because of two main reasons. First, the time series is not continuous and has many (small) intervals of no data, Fig. \ref{fig6}. When the data have gaps, the prediction is based on older data and the changes in the coordinates in between the prediction point and the training set are not considered. It is obvious, therefore, that the prediction accuracy is lower, compared to the situation in which data are continuous. The second reason is that the amplitude of changes is higher in this time series, Fig. \ref{fig7}. The more amplitude of changes the more variable the time series is, and of course the mathematical model in Eqn. \eqref{eqn1} would result in an estimate that is less accurate than that of a more uniform time series.    
Another important point that can be simply observed from tables 1, 2, and 3 is that the $Y$ component of the time series has more accurate predictions, compared to $X$ and $Z$ components. This is because in the observation files the accuracy of $Y$ component is around 1 mm, while the other two components' accuracy is 1 cm. This has an indirect effect on the prediction accuracy, because in Eqn. \eqref{eqn1} there is no explicit indication of the impact of the training data accuracy on the prediction. However, the more precise the training data, the more accurate the prediction is.

\begin{figure}[H]
	\centering
	\includegraphics[width=1\linewidth]{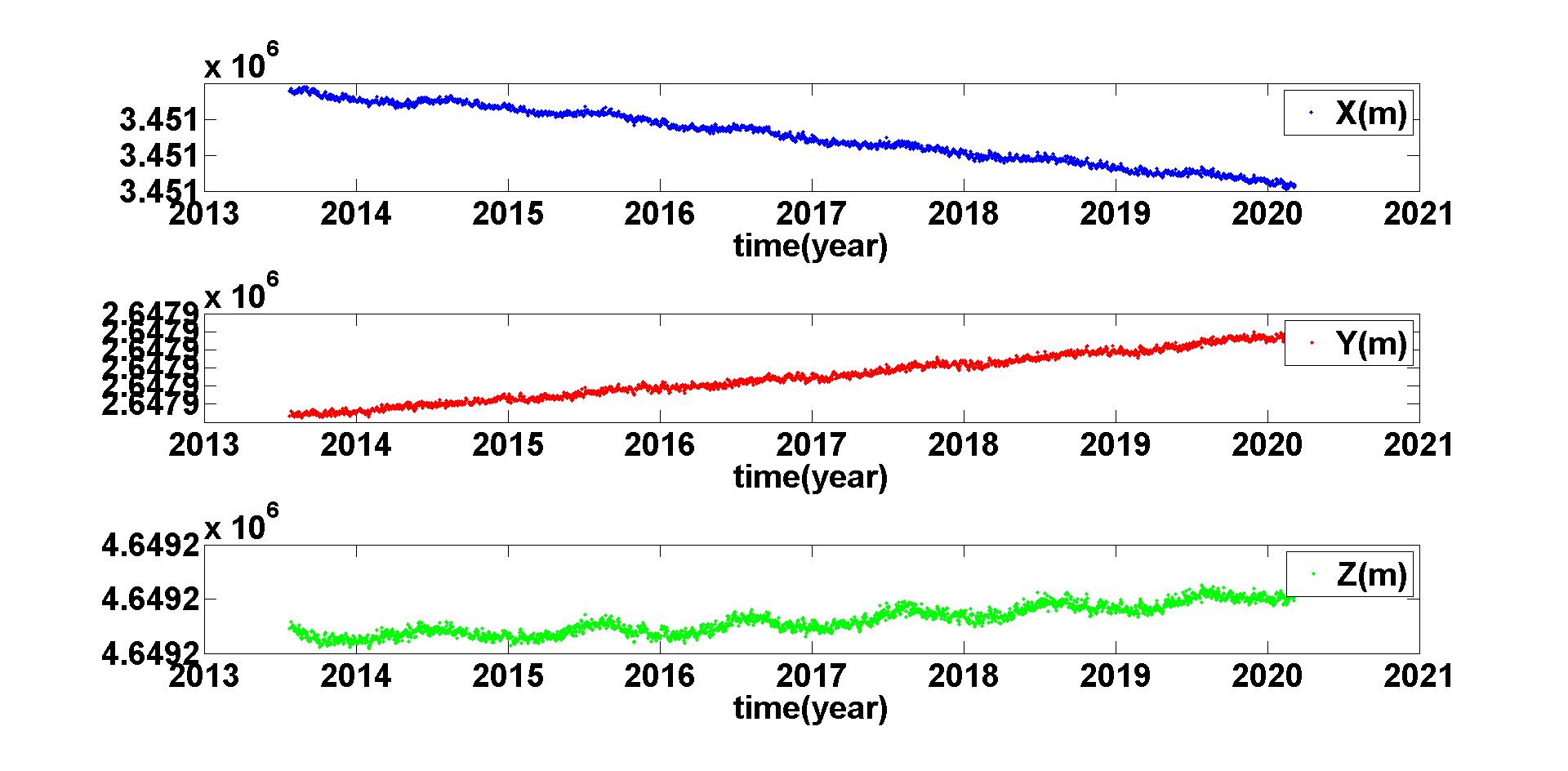}
	\caption{Continuous time series of Mariupol, Ukraine}
	\label{fig8}
\end{figure}

\begin{figure}[H]
	\centering
	\includegraphics[width=1\linewidth]{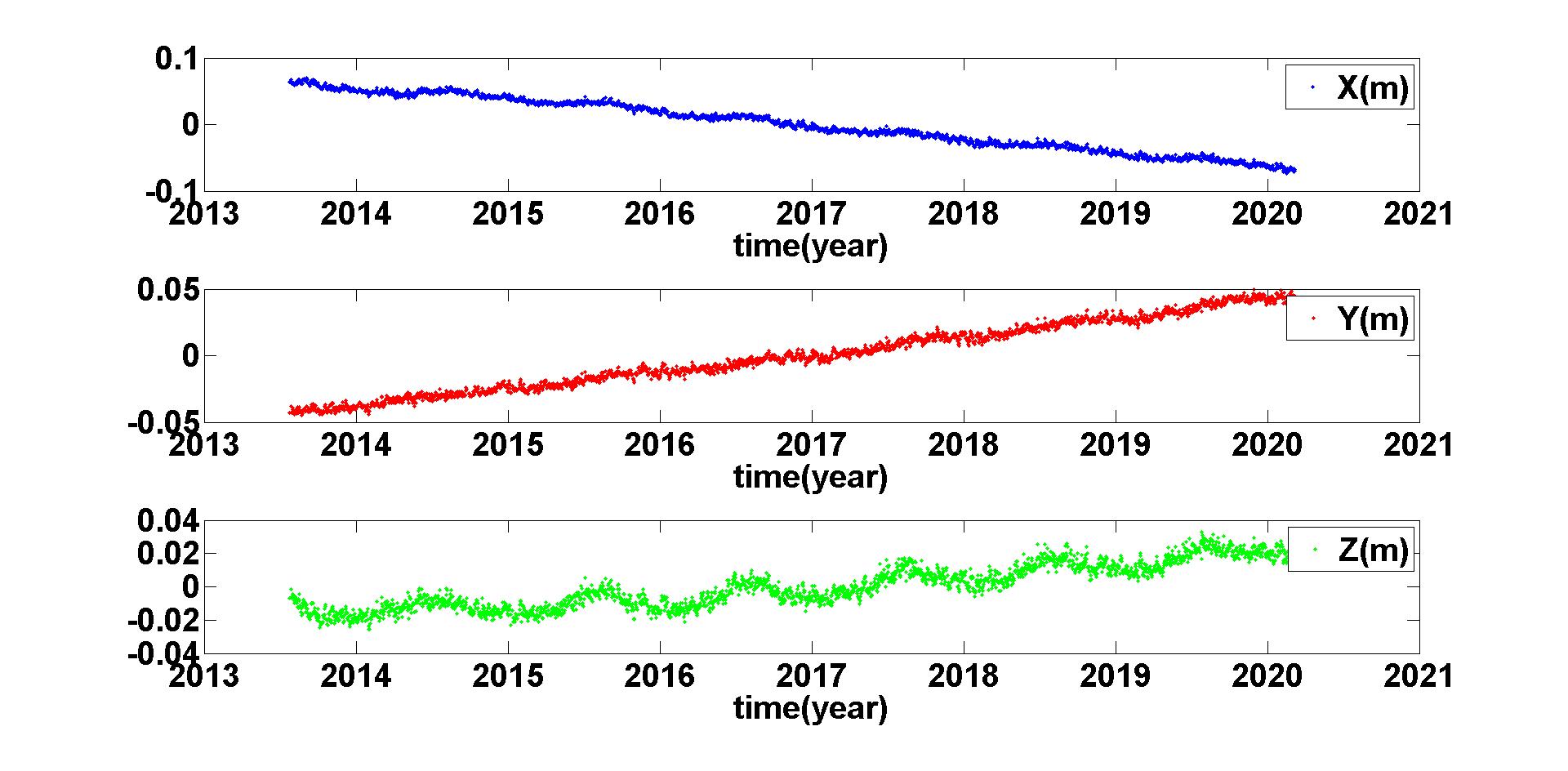}
	\caption{Amplitude of changes in the Mariupol, Ukraine time series, with respect to the mean value of components}
	\label{fig9}
\end{figure}

The most accurate prediction is for the Mariupol, Ukraine permanent GNSS station. This can be explained by the fact that first, time series is continuous-with no gaps, Fig. \ref{fig8}-and second, the time series changes very little, Fig. \ref{fig9}. In this time series, we can expect to predict the components of the time series with accuracy of around 5 centimeters.
Note also that by comparing the sMAPE of the discontinuous data with that of the continuous ones, one can roughly say the error of the prediction in discontinuous data is ten times more than that of the continuous data. It can be also understood from MAbs.

It is also interesting to investigate the effect of time series length on the overall prediction accuracy. To this end, one must refer to the tables 1, 2, and 3. In these tables, the most accurate predictions are for the Mariupol, Ukraine station. The time span for this time series in roughly 7 years, from 2013 to 2020. According to the auto-regressive nature of the GRNN, in which the predicted values become training data for the next prediction, the error in each iteration (each prediction step) is included in the next iteration. In other words, the error in GRNN is cumulative. However, this has a negligible effect on the overall accuracy of prediction, in comparison with the discontinuity and high variability effects. This can be simply understood by comparing the prediction accuracy of Kirkkonummi, Kunzak and Matera stations. Note that this observation is simply in stark contrast to the traditional methods such as adaptive numerical integrations \cite{Kiani2}, in which the cumulative error is indeed the major source of error.

\begin{figure}[H]
	\centering
	\includegraphics[width=1\linewidth]{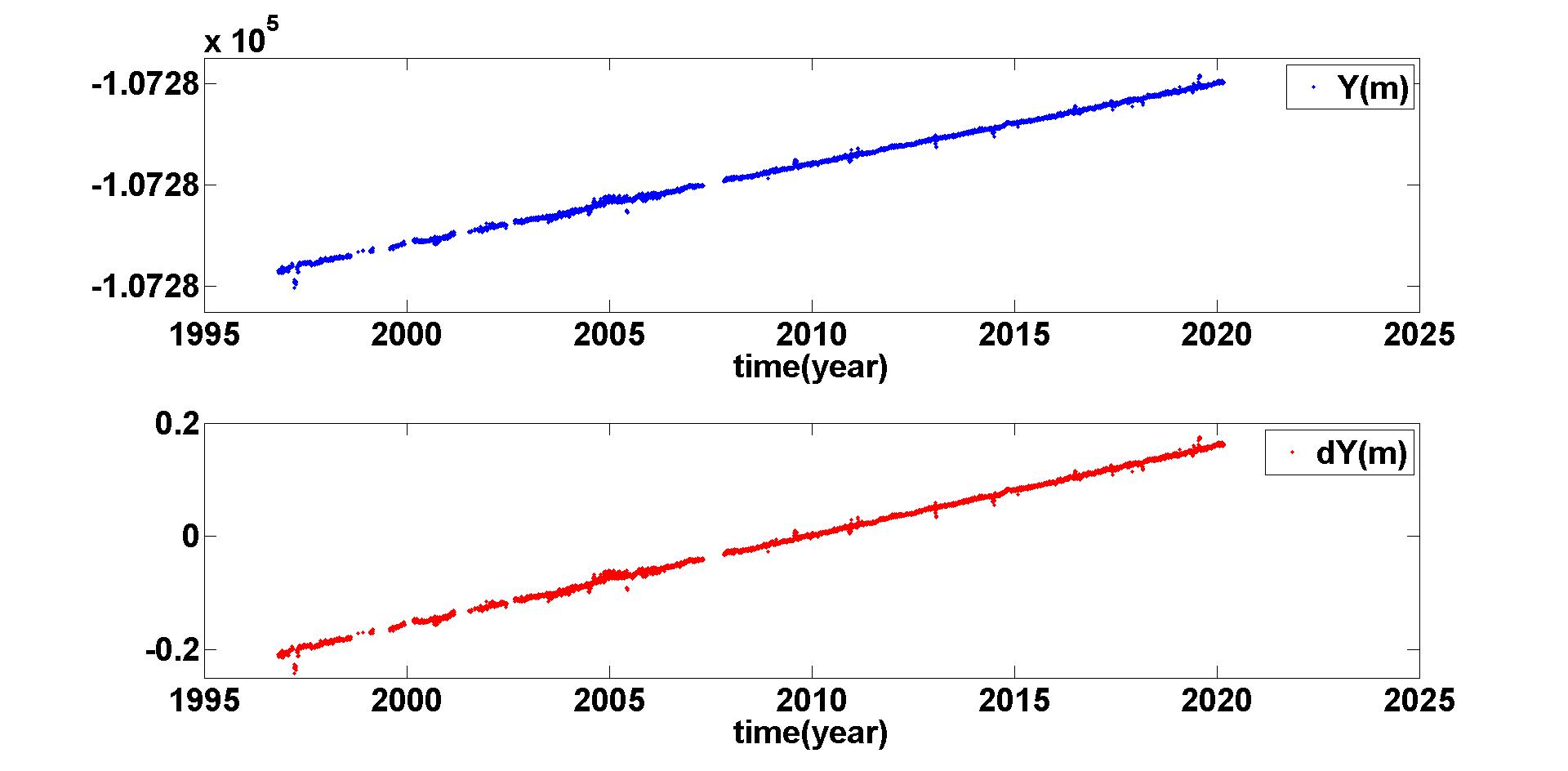}
	\caption{$Y$ component and changes, $dY$, in the Morpeth, United Kingdom time series, with respect to the mean value of $Y$}
	\label{fig10}
\end{figure}

Considering the Morpeth, United Kingdom station, one can observe that in $Y$ component it has the third most accurate results, after Mariupol and Kunzak. Even though the data are discontinuous, and the time span is relatively long (24 years) Fig. \ref{fig10}, the prediction accuracy is around 6 centimeters. This is because the time series is not variable much. It changes in an ascending trend. In 24 years, it changes almost 40 centimeters. In between, however, there are not many "unexpected" changes, meaning the changes are almost in a linear mode.       

Overall, it can be said that based on the tables 1, 2, and 3, the prediction accuracy in $X$, $Y$, and $Z$ components can reach up to 7, 3, and 5 centimeters, respectively. The worst prediction accuracy values, however, can be as high as 14, 4 and 12 meters, for $X$, $Y$, and $Z$ components, respectively.

Finally, in order to check the accuracy of GRNN machine learning algorithm against the traditional statistical methods, the Theta method \cite{Assimakopoulos} is used for predicting the 14 mentioned time series. The Theta method is based on the second discrete finite differences, which in return represent the modification of the local curvature of the time series \cite{Assimakopoulos}. This method is useful for finding the prediction values of time series with trend. Since almost all the time series we have used have trends, this method can work well in the problem. In mathematical representation, the $k$ th prediction, as in Eqn. \eqref{eqn1}, is given in the following form, based on the parameter of the method, denoted by $\theta$
\begin{equation}\label{eqn6}
\hat{y}_k=y_1+(k-1)(y_2-y_1)+\theta(\sum_{t=2}^{k-1}(k-t)(y_{t+1}-2y_t+y_{t-1})).
\end{equation}
Eqn. \eqref{eqn6} represents a line with slope $\theta$. This line is often called $\theta$-line. The slope, $\theta$, can be estimated using the following relation
\begin{equation}\label{eqn7}
\theta=\frac{12}{p(p^2-1)}\sum_{t=1}^{p}ty_t-\frac{6}{p(p-1)}\sum_{t=1}^{p}y_t.
\end{equation}
The results of applying Theta method to the time series are shown in tables 4, 5, 6. 

\begin{table}[H]
	\label{table4}
	\centering
	\caption{Prediction accuracy of the Theta method for 14 different permanent GNSS stations, $X$ component}
	\begin{tabular}{|c|c|c|c|c|c|}\hline
		\cline{1-6}station name & time span (year) & state of data & sMAPE$\textdiscount$ & StD (m) & MAbs(m)\\
		
		\cline{1-6} A Coruna, Spain & 1998-2020 & \textcolor{red}{discontinuous} & 3.732 & 13.478 & 23.353\\
		
		\cline{1-6} Ajaccio, France & 2000-2020 & \textcolor{red}{discontinuous} & 0.703 & 3.026 & 5.241\\
		
		\cline{1-6} Bacau, Romania & 2006-2020 & continuous & 0.545 & 2.515 & 4.364\\
		
		\cline{1-6} Borowa Gora, Poland & 1996-2020 & continuous & 2.243 & 5.814 & 9.876\\
		
		\cline{1-6} Svetloe, Russian Federation & 1997-2020 & continuous & 4.136 & 11.697 & 20.411\\
		
		\cline{1-6} Morpeth, United Kingdom & 1996-2020 & \textcolor{red}{discontinuous} & 2.190 & 6.571 & 11.323\\
		
		\cline{1-6} Kunzak, Czech Republic & 2005-2020 & continuous & 1.153 & 5.231 & 9.067\\
		
		\cline{1-6} Maartsbo, Sweden & 1996-2020 & continuous & 4.864 & 9.918 & 17.181\\
		
		\cline{1-6} Mariupol, Ukraine & 2013-2020 & continuous & 0.295 & 2.459 & 4.262\\
		
		\cline{1-6} Matera, Italy & 1994-2020 & continuous & 9.481 & 27.161 & 47.044\\
		
		\cline{1-6} Kirkkonummi, Finland & 2013-2020 & continuous & 0.057 & 0.597 & 1.032\\
		
		\cline{1-6} Modra-Piesok, Slovak Republic & 2007-2020 & continuous & 1.085 & 5.835 & 10.113\\
		
		\cline{1-6} Nicosia, Cyprus & 1997-2020 & \textcolor{red}{discontinuous} & 2.954 & 10.162 & 17.613\\
		
		\cline{1-6} Athens, Greece & 2006-2020 & continuous & 1.622 & 7.815 & 14.898\\
		
		\hline
	\end{tabular}
\end{table}

\begin{table}[H]
	\label{table5}
	\centering
	\caption{Prediction accuracy of the Theta method for 14 different permanent GNSS stations, $Y$ component}
	\begin{tabular}{|c|c|c|c|c|c|}\hline
		\cline{1-6}station name & time span (year) & state of data & sMAPE$\textdiscount$ & StD (m) & MAbs(m)\\
		
		\cline{1-6} A Coruna, Spain & 1998-2020 & \textcolor{red}{discontinuous} & 1.999 & 1.073 & 1.847\\
		
		\cline{1-6} Ajaccio, France & 2000-2020 & \textcolor{red}{discontinuous} & 1.503 & 0.999 & 1.728\\
		
		\cline{1-6} Bacau, Romania & 2006-2020 & continuous & 1.373 & 3.218 & 5.574\\
		
		\cline{1-6} Borowa Gora, Poland & 1996-2020 & continuous & 6.694 & 6.550 & 11.334\\
		
		\cline{1-6} Svetloe, Russian Federation & 1997-2020 & continuous & 4.873 & 8.000 & 13.761\\
		
		\cline{1-6} Morpeth, United Kingdom & 1996-2020 & \textcolor{red}{discontinuous} & 41.393 & 3.655 & 6.298\\
		
		\cline{1-6} Kunzak, Czech Republic & 2005-2020 & continuous & 0.947 &  1.165 & 2.023\\
		
		\cline{1-6} Maartsbo, Sweden & 1996-2020 & continuous & 12.109 & 7.671 & 13.288\\
		
		\cline{1-6} Mariupol, Ukraine & 2013-2020 & continuous & 0.516 & 3.299 & 5.722\\
		
		\cline{1-6} Matera, Italy & 1994-2020 & continuous & 29.206 & 25.105 & 43.489\\
		
		\cline{1-6} Kirkkonummi, Finland & 2013-2020 & continuous & 0.264 & 1.257 & 2.175\\
		
		\cline{1-6} Modra-Piesok, Slovak Republic & 2007-2020 & continuous & 1.882 & 3.151 & 5.456\\
		
		\cline{1-6} Nicosia, Cyprus & 1997-2020 & \textcolor{red}{discontinuous} & 5.412 & 12.282 & 21.274\\
		
		\cline{1-6} Athens, Greece & 2006-2020 & continuous & 1.622 & 2.148 & 4.303\\
		
		\hline
	\end{tabular}
\end{table}

\begin{table}[H]
	\label{table6}
	\centering
	\caption{Prediction accuracy of the Theta method for 14 different permanent GNSS stations, $Z$ component}
	\begin{tabular}{|c|c|c|c|c|c|}\hline
		\cline{1-6}station name & time span (year) & state of data & sMAPE$\textdiscount$ & StD (m) & MAbs(m)\\
		
		\cline{1-6} A Coruna, Spain & 1998-2020 & \textcolor{red}{discontinuous} & 1.396 & 4.787 & 8.286\\
		
		\cline{1-6} Ajaccio, France & 2000-2020 & \textcolor{red}{discontinuous} & 0.243 & 0.941 & 1.632\\
		
		\cline{1-6} Bacau, Romania & 2006-2020 & continuous & 0.337 & 1.833 & 3.174\\
		
		\cline{1-6} Borowa Gora, Poland & 1996-2020 & continuous & 2.067 & 7.352 & 12.611\\
		
		\cline{1-6} Svetloe, Russian Federation & 1997-2020 & continuous & 1.155 & 6.701 & 11.546\\
		
		\cline{1-6} Morpeth, United Kingdom & 1996-2020 & \textcolor{red}{discontinuous} & 5.081 & 21.708 & 37.581\\
		
		\cline{1-6} Kunzak, Czech Republic & 2005-2020 & continuous & 1.458 &  7.863 & 13.627\\
		
		\cline{1-6} Maartsbo, Sweden & 1996-2020 & continuous & 3.198 & 12.033 & 20.845\\
		
		\cline{1-6} Mariupol, Ukraine & 2013-2020 & continuous & 0.338 & 3.787 & 6.579\\
		
		\cline{1-6} Matera, Italy & 1994-2020 & continuous & 6.568 & 16.746 & 29.019\\
		
		\cline{1-6} Kirkkonummi, Finland & 2013-2020 & continuous & 0.092 & 1.847 & 3.196\\
		
		\cline{1-6} Modra-Piesok, Slovak Republic & 2007-2020 & continuous & 0.128 & 0.810 & 1.401\\
		
		\cline{1-6} Nicosia, Cyprus & 1997-2020 & \textcolor{red}{discontinuous} & 5.315 & 15.324 & 26.538\\
		
		\cline{1-6} Athens, Greece & 2006-2020 & continuous & 0.821 & 3.047 & 6.414\\
		
		\hline
	\end{tabular}
\end{table}
Comparing tables 1, 2, 3 and 4, 5, 6, one can understand that the results of the Theta method are more biased and more dispersed than those of the GRNN. In table 7, the relative prediction accuracy criteria of the GRNN and Theta methods are compared. The prediction accuracy criteria of the GRNN are divided by those of the Theta method.
\begin{table}[H]
	\label{table7}
	\centering
	\caption{Relative prediction accuracy and computation time of the GRNN with respect to the Theta method}
	\begin{tabular}{|c|c|}\hline
		\cline{1-2} relative criterion & value\\
		
		\cline{1-2} sMAPE, $X$ component & 1.7$\times 10^{-6}$\\
		
		\cline{1-2} sMAPE, $Y$ component & 3.2$\times 10^{-7}$\\
		
		\cline{1-2} sMAPE, $Z$ component & 1.2$\times 10^{-6}$\\
		
		\cline{1-2} StD, $X$ component & 0.02\\
		
		\cline{1-2} StD, $Y$ component & 0.004\\
		
		\cline{1-2} StD, $Z$ component & 0.01\\
		
		\cline{1-2} MAbs, $X$ component & 0.008\\
		
		\cline{1-2} MAbs, $Y$ component & 0.003\\
		
		\cline{1-2} MAbs, $Z$ component & 0.006\\
		
		\cline{1-2} computation time & 0.219\\
		
		\hline
	\end{tabular}
\end{table}
Considering table 7, one can simply deduce that the results of GRNN are at most 250 times more uniform (more accurate), and 125 times less biased with respect to their mean value. Besides, GRNN procedure is 4.6 times faster than the Theta method. Overall, therefore, it can be said the machine learning algorithm present a better approach for prediction than the traditional methods such Theta.       

\section{Conclusions and ways forward}
An applied study of the generalized regression neural network, a type of machine learning algorithm, for GNSS position time series prediction is presented. 14 permanent GNSS stations are used in this study, which are located in different countries. The symmetric mean absolute percentage error, standard deviation, and mean of absolute errors of the prediction errors are computed and compared with another independent approach, the so-called Theta method. Based on this study, the following conclusions apply to the results of prediction.

$\bullet$ The most important factors affecting the accuracy of the prediction are, in order, the availability of gaps (discontinuity) and the variability of the time series between adjacent points.

$\bullet$ The machine learning algorithm works well for both the periodic and linear (with trend) time series. However, the amplitude of changes is what controls the accuracy of prediction.

$\bullet$ The number of predictions has a negligible effect on the overall prediction accuracy, in spite of the auto-regressive nature of the machine learning algorithm.

$\bullet$ The results of the Theta method are more dispersed (higher standard deviation) and more biased (higher mean of the absolute errors), unlike the case with the M3 competition time series in \cite{Makridakis,Ahmed}. 

$\bullet$ The machine learning algorithm is faster than the Theta method.

$\bullet$ The prediction accuracy in components with higher observation accuracy is higher, even though there is no explicit indication of this in the machine learning algorithm. 

$\bullet$ In the case of using machine learning algorithm, one can expect to achieve up to a few centimeters in prediction accuracy.

The generalized regression neural network is just a member of the large class of machine learning algorithms that can be used for geodetic time series prediction. As was seen in this paper, the method presents high accuracy, up to centimeter level. However, the method is purely a mathematical model. It would be more realistic if the role of the observation accuracy and the physical conditions of the environment could be included in it. This may lead to a more accurate algorithm, capable of achieving higher accuracies, possibly up to the millimeter level. This is a good motivation to work on developing the mathematical model and formulae for this method, which would be the modified, more accurate version of the original algorithm.


\begin{thebibliography}{99990}
	\bibitem{Ahmed}Ahmed, N. K., Atiya, A. F., El Gayar, N., El-Shishiny, H.:An empirical comparison of machine learning models for time series forecasting. Econometric Reviews. 29, 594-621 (2010) doi: 10.1080/07474938.2010.481556
	
	\bibitem{Assimakopoulos}Assimakopoulos, V., Nikolopoulos, K.:The theta model: a decomposition approach to forecasting. International Journal of Forecasting. 16, 521–530, (2000)
	
	\bibitem{Awad}Awad, M.,Khanna, R.:Support vector regression. In: Efficient learning machines. Apress, Berkeley, CA. (2015) https://doi.org/10.1007/978-1-4302-5990-9-4
	
	\bibitem{Blewitt}Blewitt, G., Hammond, W., Kreemer, C.:Harnessing the GPS data explosion for interdisciplinary science. Nevada Geodetic Laboratory. (2018) doi: 10.1029/2018EO104623. 
	
	\bibitem{Boughrara}Boughrara, H., Chtourou, M., Ben Amar, C., Chen, L.:Facial expression recognition based on a mlp neural network using constructive training algorithm. Multimedia Tools and Applications. 75, 709–731 (2016)
	
	\bibitem{Dash}Dash, R., Dash, P. K.:A comparative study of radial basis function network with different basis functions for stock trend prediction. IEEE Power, Communication and Information Technology Conference. (2015)
	
	\bibitem{Du}Du, Y., Wang, W., Wang, L.:Hierarchical recurrent neural network for skeleton based action recognition. IEEE Conference on Computer Vision and Pattern Recognition. 1110-1118 (2015)
	
	\bibitem{Fazilova}Fazilova, D., Ehgamberdiev, S. H., Kuzin, S.:Application of time series modeling to a national reference frame realization. Geodesy and Geodynamics. 9, 281-287 (2018)
	
	\bibitem{Fischer}Fischer, T., Krauss, C.:Deep learning with long short-term memory networks for financial market predictions. European Journal of Operational Research. 270, 654-669 (2018)
	
	\bibitem{Hayes}Hayes, T., Usami, S., Jacobucci, R., McArdle, J. J.:Using Classification and Regression Trees (CART) and random forests to analyze attrition: Results from two simulations. Psychol Aging. 30, 911-29 (2015) doi: 10.1037/pag0000046
	
	\bibitem{Ito}Ito, C., Takahashi, H., Ohzono, M.:Estimation of convergence boundary location and velocity between tectonic plates in northern Hokkaido inferred by GNSS velocity data. Earth, Planets and Space. 71 (2019) https://doi.org/10.1186/s40623-019-1065-z
	
	\bibitem{Kiani}Kiani, M.:On the analysis of the earth’s surface vertical change by GNSS residual position time series prediction and analysis using radial basis function networks machine learning. Second International Conference on Development of Materials Engineering Technology, Mining, and Geology, Iran. (2020) 
	
	\bibitem{Kiani2}Kiani Shahvandi, M.:Numerical solution of ordinary differential equations in geodetic science using adaptive Gauss numerical integration method. Acta Geod Geophys (2020). https://doi.org/10.1007/s40328-020-00293-6	
	
	\bibitem{Lauret}Lauret, P., Fock, E., Randrianarivony, R. N., Manicom-Ramsamya, J. F.:Bayesian neural network approach to short time load forecasting. Energy Conversion and Management. 49, 1156-1166 (2008)
	
	\bibitem{Li}Li, H. Z., Guo, S., Li, C. J., Sun, J. Q.:A hybrid annual power load forecasting model based on generalized regression neural network with fruit fly optimization algorithm. Knowledge-Based Systems. 37, 378-387 (2013)
	
	\bibitem{Liu}Liu, G., Ai, J.:Analysis based on RBF neural network for effect of artificial flow field on dissolved oxygen uniform distribution. Advances in Intelligent Systems Research. 115, International Conference on Electronic Industry and Automation. (2017)
	
	\bibitem{Makridakis}Makridakis, S., Spiliotis, E., Assimakopoulos, V.:statistical and machine learning forecasting methods: concerns and ways forward. PLoS ONE. 13, (2018) https://doi.org/10.1371/journal.pone.0194889
	
	\bibitem{Mathur}Mathur, N., Glesk, I., Buis, A.:Comparison of adaptive neuro-fuzzy inference system (ANFIS) and Gaussian processes for machine learning (GPML) algorithms for the prediction of skin temperature in lower limb prostheses. Medical Engineering and Physics. 38, 1083-1089 (2016)
	
	\bibitem{Nadaraya}Nadaraya, E. A.:On estimating regression. Theory of Probability and Its Applications. 10, 186–190 (1964)
	
	\bibitem{Orr}Orr, M. J. L.:A Introduction to radial basis function networks. Centre for Cognitive Science, Scotland (1996)
	
	\bibitem{Rajini}Rajini, N. H., Bhavani, R.:Classification of MRI brain images using k-nearest neighbor and artificial neural network. IEEE International Conference on Recent Trends in Information Technology, India. (2011) doi: 10.1109/ICRTIT.2011.5972341
	
	\bibitem{Watson}Watson, G. S.:Smooth regression analysis. Sankhy Series A. 26, 359–372 (1964)
	
	\bibitem{Zhang}Zhang, M. L.:ML-RBF: RBF neural networks for multi-label learning. Neural Processing Letters. 29, 61–74 (2009)
\end{thebibliography}
\end{document}